\documentclass[aps,twocolumn,showpacs]{revtex4}
\usepackage{bm}
\usepackage{graphicx}
\usepackage{epsfig}
\begin{document}

\title{Spin orientation of a two-dimensional electron gas by a high-frequency
electric field}
\author{S.\,A.\,Tarasenko}
\affiliation{A.F.~Ioffe Physico-Technical Institute of the Russian
Academy of Sciences, 194021 St.~Petersburg, Russia}
\begin{abstract}
Coupling of spin states and space motion of conduction electrons
due to spin-orbit interaction opens up possibilities for
manipulation of the electron spins by electrical means. It is
shown here that spin orientation of a two-dimensional electron gas
can be achieved by excitation of the carriers with a linearly
polarized high-frequency electric field. In (001)-grown quantum
well structures excitation with in-plane ac electric field induces
orientation of the electron spins along the quantum well normal,
with the spin sign and the magnitude depending on the field
polarization.
\end{abstract}


\maketitle

\section{Introduction}
Spin dynamics of charge carriers and, particularly, possibilities
of manipulating electron spins in semiconductor structures is
attracting a great deal of attention~\cite{spintronics}. Much
effort in this research area is directed towards the development
of efficient methods of injection and detection of spin-polarized
carriers by electrical means. In particular, it has been shown
that in gyrotropic semiconductor structures, i.e. structures of
symmetry classes which allow spin-orbit splitting of the spectrum
linear in the wave vector, spins of the free carriers can be
oriented by an electric current
flow~\cite{Ivchenko,Vasko,Aronov,Edelstein}. This current-induced
spin polarization has been observed in bulk
tellurium~\cite{Vorob'ev} and recently in strained bulk
InGaAs~\cite{Kato} and GaAlAs quantum well (QW)
structures~\cite{Silov,Ganichev}. Microscopically, the effect
represents a current-induced selective occupation of the spin
subbands, which are split in $\bm k$-space due to the spin-orbit
interaction. Dynamics of the spin polarization is governed by the
spin relaxation time that determines the rate of carrier
redistribution between the spin subbands. The spin polarization
induced by ac electric field oscillates at the field frequency,
its amplitude being determined by the ratio between the field
frequency and the spin relaxation rate. Under application of ac
electric field of frequency higher than the spin relaxation rate,
the spin polarization vanishes. On the other hand, as is known,
perturbation of the electron gas with ac electric field causes an
absorption of the field by the free carriers. In systems with
spin-orbit interaction this process may also be
spin-dependent~\cite{Rashba}.

In this paper we show that perturbation of a two-dimensional (2D)
electron gas with a linearly polarized high-frequency electric
field induces spin orientation of the carriers. The direction of
the spin orientation is determined by the field polarization and
the explicit form of the spin-orbit interaction. In particular,
excitation with in-plane ac electric field in conventional
(001)-grown QW structures leads to orientation of the electron
spins along the QW normal, its sign and magnitude depending on the
field polarization with respect to the crystallographic axes. From
the phenomenological symmetry analysis, the effect under study is
similar to the interband optical orientation of electron spins by
linearly polarized light~\cite{Tarasenko}. However, in contrast to
the direct optical transitions from the valence to the conduction
band, here only one type of carriers, namely, the conduction
electrons, is excited by the electromagnetic field. Thus, the
microscopic mechanism of the spin orientation of 2D electron gas
proposed in the present paper differs from that based on the
selection rules for interband optical transitions.

\section{Microscopic model}

Absorption of a high-frequency electric field by free carriers, or
Drude-like absorption, occurs in doped semiconductor structures
and is always accompanied by electron scattering from acoustic or
optical phonons, static defects, etc., because of the need for
energy and momentum conservation. In systems with a spin-orbit
interaction, processes involving change of the particle wave
vector are spin-dependent. In particular, the matrix element of
electron scattering by static defects or phonons
$V_{\bm{k}'\bm{k}}$ in QW structures contains, in addition to the
main contribution $V_0$, an asymmetric spin-dependent
term~\cite{Belinicher,Averkiev,monopolar}
\begin{equation}\label{V_asym}
V_{\bm{k}'\bm{k}} = V_0 + \sum_{\alpha\beta} V_{\alpha\beta} \,
\sigma_{\alpha} (k_{\beta}+k'_{\beta}) \:,
\end{equation}
where $\bm{k}$ and $\bm{k}'$ are the initial and the scattered
in-plane wave vectors, respectively, and $\sigma_{\alpha}$
($\alpha=x,y,z$) are the Pauli matrices. Microscopically, this
contribution originates from structural or bulk inversion
asymmetry similar to $\bm{k}$-linear Rashba and Dresselhaus spin
splitting of the electron subbands in QWs grown from
zinc-blende-type compounds. Due to the spin-dependent asymmetry of
the scattering, electrons photoexcited from the subband bottom are
scattered in preferred directions depending on their spin
states~\cite{spinflux,Belkov}. This concept is illustrated in
Fig.~1(a), where the free-carrier absorption is shown as a
combined two-stage process involving electron-photon interaction
(vertical solid lines) and electron scattering (dashed horizontal
lines). The scattering asymmetry is shown by thick and thin dashed
lines: electrons with the spins $+1/2$ and $-1/2$ are scattered
predominantly into the states with $k_x>0$ and $k_x<0$,
respectively. Note, that such an electron distribution represents
a pure spin current~\cite{spinflux}, i.e. a state where particles
with opposite spins flow in opposite directions, while the average
electron spin remains zero. A net spin orientation of the electron
gas appears as a result of the the subsequent spin dynamics of the
carriers. The spin dynamics of the conduction electrons is known
to be governed by spin-orbit coupling that may be considered as an
effective magnetic field that acts on the electron
spins~\cite{DP}. The corresponding Hamiltonian of the spin-orbit
interaction has the form
\begin{figure}[t]
\leavevmode \epsfxsize=0.95\linewidth
\centering{\epsfbox{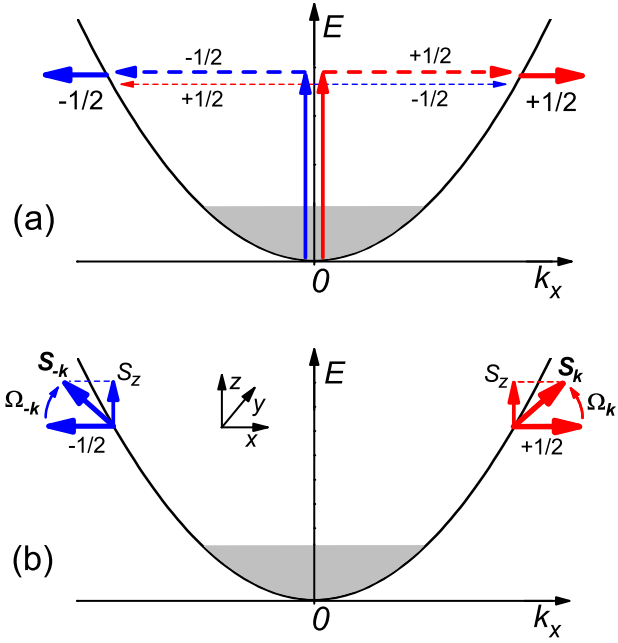}} \caption{(Color online).
Microscopic origin of the spin orientation of 2D electron gas by a
high-frequency electric field. (a) Asymmetry of scattering under
photoexcitation followed by (b) spin precession in the effective
magnetic field leads to appearance of the average electron spin.}
\vspace{-0.5cm}
\end{figure}
\begin{equation}
H_{so}= \frac{\hbar}{2} (\bm{\Omega}_{\bm{k}} \cdot \bm{\sigma})
\:,
\end{equation}
which is similar to the Hamiltonian describing Zeeman splitting of
the spectrum in the external magnetic field. Of special note is
that the direction and strength of the effective field and,
correspondingly, the direction and strength of the Larmor
frequency of the field, $\bm{\Omega}_{\bm{k}}$, depend on the
electron wave vector. In this effective magnetic field the spins
of the photoexcited carriers, originally directed according to the
spin-dependent scattering, precess as shown in Fig.~1(b).
Electrons with the initial spin $+1/2$ and the wave vector $k_x>0$
are affected by the field with the Larmor frequency
$\bm{\Omega}_{\bm{k}}$, while carriers with the opposite spin,
$-1/2$, and the opposite wave vector are affected by the field
with the frequency $\bm{\Omega}_{-\bm k}$. Since the effective
magnetic field induced by spin-orbit coupling is an odd function
of the wave vector, $\bm{\Omega}_{-\bm k}=-\bm{\Omega}_{\bm
k}$~\cite{DP}, the rotation axes for carriers with the initial
spins $\pm 1/2$ are opposite. As a consequence of this precession
the spin component $S_z>0$ appears for the carriers with both
positive and negative $k_x$ as is shown in Fig.~1(b), yielding a
net spin polarization of the electron gas. The spin generation
rate under steady-state excitation is determined by the average
angle of the spin rotation in the effective magnetic field,
similarly to the appearance of a perpendicular spin component in
the Hanle effect.

\section{Theory}

Indirect optical transitions are treated in perturbation theory as
second-order processes involving virtual intermediate states. The
compound matrix element of this kind of the transition with the
initial $|s\,\bm{k}\rangle$ and final $|s'\,\bm{k}'\rangle$ states
in the electron subband $e1$ has the form (see, e.g.,
Ref.~\onlinecite{monopolar})
\begin{equation}\label{M_gen}
M_{s'\bm{k}',s\bm{k}} = \sum_{j} \left(
\frac{V_{e1s'\bm{k}',j\bm{k}}
R_{j\bm{k},e1s\bm{k}}}{E_{e1\bm{k}}-E_{j\bm{k}}+\hbar\omega} +
\frac{R_{e1s'\bm{k}',j\bm{k}'} V_{j\bm{k}',e1s\bm{k}}}
{E_{e1\bm{k}}-E_{j\bm{k}'}} \right) \:.
\end{equation}
Here $s$ and $s'$ are the spin indices, $j$ denotes the subband of
an intermediate state; $E_{e1\bm{k}}$, $E_{e1\bm{k}'}$ and
$E_{j\bm{k}}$ are the electron energies in the initial, final and
intermediate states, respectively; $V_{e1s'\bm{k}',j\bm{k}}$ is
the matrix element of electron scattering, $R_{j\bm{k},e1s\bm{k}}$
is the matrix element of electron interaction with the
electromagnetic field, and $\omega$ is the field frequency.
Spin-orbit splitting of the subbands $e1$ and $j$, odd in the wave
vector, is neglected in Eq.~(\ref{M_gen}), since it leads to no
essential contribution to the pure spin current induced by
free-carrier absorption.

A dominant contribution to absorption of a high-frequency electric
field in QWs is introduced by processes with intermediate states
in the same subband, $e1$. Taking into account the matrix element
of electron scattering within the subband $e1$ in the form of
Eq.~(\ref{V_asym}), one derives two contributions to the matrix
element of the indirect optical transitions
\begin{equation}\label{M0}
M_{\bm{k}'\bm{k}}^{(0)} = \frac{eA}{c\omega m^*}\,
\bm{e}\cdot(\bm{k}-\bm{k}') \,V_0 \:,
\end{equation}
\begin{equation}\label{M1a}
M_{\bm{k}'\bm{k}}^{(1\mathrm{a})} = \frac{eA}{c\omega m^*}\,
\bm{e}\cdot(\bm{k}-\bm{k}') \sum_{\alpha\beta} V_{\alpha\beta}
\sigma_{\alpha} (k_{\beta} + k_{\beta}') \:,
\end{equation}
where $e$ is the elementary charge, $c$ is the light velocity,
$m^*$ is the effective electron mass, $\bm{A}=A\bm{e}$ is the
vector potential of the field, $\bm{e}$ is the (unit) polarization
vector. We assume the electromagnetic field to be linearly
polarized and, therefore, consider the polarization vector
$\bm{e}$ to be real. The matrix element~(\ref{M0}) determines the
QW absorption coefficient, but does not contribute to spin
phenomena alone, being independent of the electron spin state. On
the contrary, the term given by Eq.~(\ref{M1a}) gives rise to
spin-related effects under indirect optical transitions because of
the spin-dependent scattering. In general, a dependence of the
amplitude of electron scattering on the spin can be derived if one
considers $\bm{k} \cdot \bm{p}$ admixture of the valence-band
states to the conduction-band wave functions and takes into
account the spin-orbit splitting of the valence band. To first
order in the in-plane wave vector, the contribution to the matrix
element of electron scattering by short-range static defects has
the form
\begin{equation}\label{V1}
V_{s'\bm{k}', s\bm{k}}^{(1)}= \frac{\hbar}{m_0} \sum_{j \neq e1}
\frac{V_{e1 s', j} \: \bm{k} \cdot \bm{p}_{j, e1s} + \bm{k}' \cdot
\bm{p}_{e1 s', j} \, V_{j, e1s}}{E_{e1} - E_j} \:,
\end{equation}
where $m_0$ is the free-electron mass, $E_{e1}$ and $E_j$ are the
subband energies at $\bm{k}=0$, and $\bm{p}_{j,e1s}$ is the matrix
element of the momentum operator. This contribution leads to
spin-dependent terms in the matrix element of electron scattering
which may be rewritten in the form of Eq.~(\ref{V_asym}), with the
coefficients $V_{\alpha\beta}$ given by
\begin{equation}\label{V_ab}
V_{\alpha\beta} = \frac{\hbar}{2m_0} \sum_{ss'=\pm1/2 }
\sum_{j\neq e1} (\sigma_{\alpha})_{ss'} \frac{(p_{\beta})_{e1s',j}
\: V_{j, e1s} \: } {E_{e1} - E_j} \:.
\end{equation}
We note that, in accordance with the time inversion symmetry, the
coefficients of the terms $\sigma_{\alpha}k_{\beta}$ and
$\sigma_{\alpha}k_{\beta}'$ in the matrix element of electron
scattering appear to be real in the appropriate basis and
coincide, as presented in Eq.~(\ref{V_asym}). This also follows
from Eq.~(\ref{V1}), taking into account that all electron and
hole states $j$ are two-fold degenerate at $\bm{k}=0$. The states
$j$ and $j^{\,\prime}$ corresponding to the same energy $E_j$ are
interconnected by the time inversion operator. Contributions to
$V_{\alpha\beta}$ from the states $j$ and $j^{\,\prime}$ are
complex conjugate fulfilling the requirements imposed on the
matrix element of electron scattering by the time inversion
symmetry.

Together with the spin-dependent contribution to the indirect
optical transitions $M_{\bm{k}'\bm{k}}^{(1\mathrm{a})}$, a
comparable term comes from the processes with virtual intermediate
states in other, e.g. valence, subbands. This term is given by
\begin{equation}
M_{s'\bm{k}',s\bm{k}}^{(1\mathrm{b})} = \frac{eA}{cm_0}
\sum_{j\neq e1} \frac{V_{e1 s', j} \: \bm{e} \cdot \bm{p}_{j, e1s}
+ \bm{e} \cdot \bm{p}_{e1s', j} \, V_{j, e1s}}{E_{e1} - E_j} \:.
\end{equation}
As is shown, it is governed by the matrix elements of interband
coupling responsible for the spin-dependent scattering,
Eq.~(\ref{V1}). Therefore, its spin-dependent part may be
expressed in terms of the coefficients $V_{\alpha\beta}$ as
follows
\begin{equation}\label{M1b}
M_{\bm{k}'\bm{k}}^{(1\mathrm{b})} = 2\frac{eA}{c\hbar}
\sum_{\alpha\beta} V_{\alpha\beta} \,\sigma_{\alpha} e_{\beta} \:.
\end{equation}

To summarize, to first order in spin-orbit interaction the matrix
element of the indirect optical transitions accompanied by
electron scattering from short-range potentials may be presented
as a sum of three contributions
\begin{equation}
M_{\bm{k}'\bm{k}} = M_{\bm{k}'\bm{k}}^{(0)} +
M_{\bm{k}'\bm{k}}^{(1\mathrm{a})} +
M_{\bm{k}'\bm{k}}^{(1\mathrm{b})} \:,
\end{equation}
with the terms given by Eqs.~(\ref{M0},\ref{M1a},\ref{M1b}),
respectively.

We assume the frequency of the ac electric field to exceed the
reciprocal relaxation times of the carriers. Then, the interaction
of the electrons with the electromagnetic field may be treated
quantum-mechanically, assuming that the field induces indirect
optical transitions. Neglecting the spin-orbit splitting of the
electron subband, the spin matrix of the carrier photogeneration
has the form (see, e.g., Ref.~\onlinecite{monopolar})
\begin{eqnarray}\label{G}
G = \frac{2 \pi}{\hbar} \sum_{\bm{k}'}
(f_{\bm{k}'}^{(0)}-f_{\bm{k}}^{(0)}) [M_{\bm{k}\bm{k}'}
M_{\bm{k}\bm{k}'}^{\dag}
\delta(\varepsilon_{\bm{k}'}-\varepsilon_{\bm{k}}+\hbar\omega) \nonumber \\
+ M_{\bm{k}'\bm{k}}^{\dag} M_{\bm{k}'\bm{k}}
\delta(\varepsilon_{\bm{k}'}-\varepsilon_{\bm{k}}-\hbar\omega)]
\:, \;\;\;
\end{eqnarray}
where $f_{\bm{k}}^{(0)}$ is the equilibrium carrier distribution
function and $\varepsilon_{\bm{k}}=\hbar^2k^2/2m^*$ is the kinetic
electron energy for the in-plane motion.

As is shown above, the absorption of the high-frequency electric
field by free carriers results in an asymmetrical spin-dependent
distribution where particles with the opposite spins flow in the
opposite directions. Such a photoexcitation asymmetry induced by
the spin-dependent scattering does not generally correspond to the
eigenstate of the spin-orbit coupling in the subband. The
subsequent rotation of the electron spins in the spin-orbit
interaction-induced effective magnetic field leads to appearance
of a net spin polarization of the electron gas. We assume the
relaxation time of the pure spin current $\tau_e$ to be shorter
than the Larmor precession period, $\Omega_{\bm{k}} \tau_e \ll 1$.
Then, in the steady-state regime, the spin generation rate is
given by~\cite{Tarasenko}
\begin{equation}\label{S_gen}
\dot{\bm{S}} = \sum_{\bm{k}} \tau_e [\bm{\Omega}_{\bm{k}} \times
\bm{g}_{\bm{k}}] \:,
\end{equation}
where the Larmor frequency corresponding to effective magnetic
field in QWs, $\bm{\Omega}_{\bm{k}}$, is a linear function of the
electron wave vector
\begin{equation}
\Omega_{\bm{k},\alpha}= \frac{2}{\hbar} \sum_{\beta}
\gamma_{\alpha\beta} k_{\beta} \:,
\end{equation}
and $\bm{g}_{\bm{k}}$ is the spin part of the photogeneration
matrix~(\ref{G}), $\bm{g}_{\bm{k}}=\mathrm{Tr}(\bm{\sigma} G)/2$.

Finally, assuming that the high-frequency electric field is
polarized in the QW plane, one derives the electron spin
generation rate
\begin{equation}\label{S}
\dot{S}_{\alpha} = \sum_{\beta\lambda\nu\mu}
\epsilon_{\alpha\beta\lambda}
 (\gamma_{\beta\nu} -
2e_{\nu}e_{\mu}\gamma_{\beta\mu}) \frac{V_{\lambda\nu}}{V_0}
\frac{\tau_e m^*}{\hbar^3} I\eta \:.
\end{equation}
Here $\epsilon_{\alpha\beta\lambda}$ is the third-rank
antisymmetric (Levy-Civita) tensor, $I$ is the light intensity
related to the vector potential of the electromagnetic field by
$I=A^2\omega^2 n_{\omega}/(2\pi c)$, $n_{\omega}$ is the
refractive index of the medium, and $\eta$ is the QW light
absorbance in this spectral range. The latter is given by
\begin{equation}
\eta = \frac{2\pi \tilde{\alpha}}{n_{\omega}} \frac{V_0^2
N_d}{(\hbar\omega)^2} N_e \kappa \:,
\end{equation}
where $\tilde{\alpha}=e^2/\hbar c$ is the fine-structure constant,
$N_e$ is the electron concentration, $N_d$ is the sheet density of
defects, and $\kappa$ is a dimensionless parameter that depends on
the carrier distribution,
\[
\kappa = \left. \int (1+2\varepsilon/\hbar\omega)
(f_{\varepsilon}^{(0)}-f_{\varepsilon+\hbar\omega}^{(0)})\,
d\varepsilon \right/ \int f_{\varepsilon}^{(0)} d\varepsilon \:,
\]
and is equal to $1$ and $2$ for the limiting cases $\hbar\omega
\gg \bar{\varepsilon}$ and $\hbar\omega \ll \bar{\varepsilon}$,
respectively, with $\bar{\varepsilon}$ being the mean electron
kinetic energy.

In the case of elastic scattering by static defects, the
relaxation time $\tau_e$ coincides with the conventional momentum
relaxation time and is governed by the same matrix element of
scattering $V_0$ that determines the QW absorbance,
\[
\tau_e^{-1}=V_0^2 N_d \,m^*/\hbar^3 \:.
\]
Assuming this, the equation for the spin generation rate~(\ref{S})
can be simplified to the following expression
\begin{equation}\label{S2}
\dot{S}_{\alpha} = \frac{2\pi \tilde{\alpha}}{n_{\omega}}
\sum_{\beta\lambda\nu\mu} \epsilon_{\alpha\beta\lambda}
(\gamma_{\beta\nu} - 2e_{\nu}e_{\mu}\gamma_{\beta\mu})
\frac{V_{\lambda\nu}}{V_0 } \frac{N_e \kappa}{(\hbar\omega)^2} \,I
\:.
\end{equation}
It is independent of the structure mobility and determined by the
ratio of the spin-dependent to the spin-independent parts of the
scattering amplitude, $V_{\lambda\nu}/V_0$, as well as by the
constants of the subband splitting, $\gamma_{\beta\nu}$. However,
it should be noted, that electron-electron collisions, which do
not affect the mobility, can control the spin dynamics and spin
transport of the conduction electrons and decrease the time
$\tau_e$~\cite{Brand,Weber}.

\section{(001)-grown quantum wells}

The direction of the spin orientation induced by ac electric field
depends on the field polarization and the explicit form of the
spin-dependent scattering and the spectrum spin splitting. The
latter is governed by the QW symmetry and can be varied. In
(001)-grown QWs based on zinc-blende-lattice semiconductors, there
are two types of $\bm k$-linear contributions to the spin-orbit
splitting of the conduction subbands. First, a contribution can
originate from the lack of an inversion center in the bulk
compositional semiconductors and/or from the anisotropy of
chemical bonds at the interfaces (so-called Dresselhaus term).
Second, $\bm k$-linear spin-orbit splitting can be induced by the
heterostructure asymmetry unrelated to the crystal lattice (Rashba
term). Similarly to the spectrum splitting, Dresselhaus-type and
Rashba-type contributions to the spin-dependent part of the
electron scattering amplitude can be distinguished.
Correspondingly, non-zero components of the tensors
$\gamma_{\alpha\beta}$ and $V_{\alpha\beta}$ are expressed in
terms of the Dresselhaus and the Rashba contributions as follows
\begin{eqnarray}\label{R_D}
\gamma_{x'y'} &=& \gamma_D + \gamma_R \:,\; \gamma_{y'x'} =
\gamma_D - \gamma_R \:,\\
V_{x'y'} &=& V_D + V_R \:,\; V_{y'x'} = V_D - V_R \:, \nonumber
\end{eqnarray}
where $x'\parallel[1\bar{1}0]$ and $y'\parallel[110]$ are the axes
in the QW plane, and $z\parallel[001]$ is the structure normal.
Experimental data indicate that the Dresselhaus and Rashba
contributions to the $\bm k$-linear subband splitting in real 2D
structures are comparable (see references in
Ref.~\onlinecite{Rashba}). Furthermore, the ratio between these
terms may be tuned, varying the heteropotential asymmetry with an
external gate voltage.

Substituting Eq.~(\ref{R_D}) into~(\ref{S}), one obtains that
excitation with the in-plane linearly polarized ac electric field
in (001)-grown QWs leads to orientation of the electron spins
along the QW normal, with the spin generation rate being
\begin{equation}\label{S_z}
\dot{S}_z = 4 e_{x'}e_{y'} (\gamma_{D}V_{R} - \gamma_{R}V_{D})
\frac{m^* \tau_e}{V_0 \hbar^3} I \eta \:.
\end{equation}

The spin orientation~(\ref{S_z}) depends on the polarization of
the ac electric field, that provides additional experimental means
to observe the effect under study. Indeed, the spin generation
$\dot{S}_z$ has opposite sign for the field polarized along the
$[100]$ and $[010]$ crystallographic axes and vanishes for the
field polarized along the $[1\bar{1}0]$ or $[110]$ axes. In
general, the dependence of the spin orientation on the field
polarization is given by $e_{x'}e_{y'}\propto\sin 2\varphi$, where
$\varphi$ is the angle between the polarization vector $\bm e$ and
the $[1\bar{1}0]$ axis.

The spin generation rate given by Eq.~(\ref{S_z}) requires both
the Rashba and the Dresselhaus contributions to the spin-dependent
scattering and the subband splitting. In particular, the spin
orientation vanishes, if the spin effects are determined only by
the Dresselhaus term, as happens in the symmetrical (001)-grown
QWs, or if they are related only to the Rashba contribution, as
can be the case of asymmetrical structures grown of
centrosymmetrical compounds.

Finally, we present an estimation for the efficiency of the
electron spin orientation by the high-frequency electric field.
The spin orientation efficiency may be consided as a ratio of the
generated spins to the total energy absorbed in the structure,
$\dot{S}/I\eta$. Following Eq.~(\ref{S_z}) one estimates the
efficiency as $\dot{S}/I\eta \sim 1$~eV$^{-1}$ for the structure
parameters: $\gamma_{\alpha\beta}/\hbar \sim 10^{5}$~cm/s,
$V_{\alpha\beta}/V_0 \sim 10^{-8}$~cm, and $\tau_e \sim
10^{-11}$~s. This value corresponds with the efficiency of the
conventional interband optical orientation by circularly polarized
light, where photons of the energy comparable to the band gap,
$E_g \sim 1$~eV, excite the spin-polarized carriers. The spin
orientation by ac electric field has an advantage that only one
type of carriers, here electrons, are involved in excitation,
allowing to consider it as a kind of spin injection. Recent
progress in optical spectroscopy of spin polarization by means of
the magneto-optical Faraday and Kerr
rotation~\cite{Kato,Ganichev,Crooker} allows one to expect that
the effect discussed in this paper is observable at present.

In conclusion, it is shown that the absorption of linearly
polarized high-frequency electric field by two-dimensional
electron gas leads to spin orientation of the carriers. The
direction and sign of the spin orientation are determined by the
polarization of ac electric field as well as by the structure
symmetry.

This work was supported by the RFBR, INTAS, programs of the RAS,
Russian Science Support Foundation, President Grant for young
scientists, and Foundation ``Dynasty'' - ICFPM.

\end{document}